# SmNiO$_3$/NdNiO$_3$ thin film multilayers

C. Girardot[1,2], S. Pignard[1], F. Weiss[1], J. Kreisel[1,*]

[1] Laboratoire Matériaux et Génie Physique, CNRS, Grenoble Institute of Technology, Minatec, 3, parvis Louis Néel, 38016 Grenoble, France

[2] Schneider Electric Industries S.A.S., 37 Quai Paul Louis Merlin, 38050 Grenoble Cedex 9 France

**Abstract**

Rare earth nickelates *RE*NiO$_3$ which attract interest due to their sharp metal-insulator phase transition, are instable in bulk form due to the necessity of an important oxygen pressure to stabilize Ni in its 3+ state of oxidation. Here, we report the stabilization of rare earth nickelates in [(SmNiO$_3$)$_t$/(NdNiO$_3$)$_t$]$_n$ thin film multilayers, *t* being the thickness of layers alternated *n* times. Both bilayers and multilayers have been deposited by Metal-Organic Chemical Vapour Deposition. The multilayer structure and the presence of the metastable phases SmNiO$_3$ and NdNiO$_3$ are evidenced from by X-ray and Raman scattering. Electric measurements of a bilayer structure further support the structural quality of the embedded rare earth nickelate layers.

* Corresponding author:  *jens.kreisel@grenoble-inp.fr*

ABO$_3$ perovskite-type thin films represent an area of increasing research interest in the field of functional oxides. A considerable amount of current research concentrates on thin film *multi*layers [1-3], which are defined as a sequence of thin film layers described as [(ABO$_3$)$_{t1}$/(A'B'O$_3$)$_{t2}$]$_n$ with $t_1$ and $t_2$ being the thicknesses of individual layers alternated $n$ times. Multilayers are often also called superstructures, superlattices or heterostructures; layers with $n$ = 1 are called bilayers.

The interest into multilayers are exemplified by three main issues (*i*) The properties of a multilayer of two or more materials may be superior to the parent materials from which they have been fabricated, as illustrated by the reported enhancement of the dielectric constant in BaTiO$_3$-SrTiO$_3$-based multilayers[4] (*ii*) The fine control of the interfaces by modern deposition techniques ("interface engineering") may lead to new unexpected properties as illustrated by intruiging conducting electron systems at the interface between insulating SrTiO$_3$ and LaAlO$_3$ layers [5,6], (*iii*) Multilayers are multifunctional materials *par excellence*, because they combine the different properties of the individual layers. An example of current interest is the combination of ferromagnetic and ferroelectric layers which lead to a multiferroic material[7,8].

For single-layer films, it is well accepted that metastable materials may be accessed through epitaxial stabilization [9-11], which enlarges the spectrum of new functional materials. However, it is less explored if such stabilization can be achieved and maintained in multilayers where each of the individual layers is in its bulk form instable.

In this study we present an investigation of multilayers based on perovskite-type *RE*NiO$_3$ (*RE* = Rare Earth), a material which is instable in bulk form due to the necessity of an important oxygen pressure to stabilize Ni in its 3+ state of oxidation.[12] The feasibility of stabilizing single layer *RE*NiO$_3$ films by epitaxial strain up to a critical thickness has been demonstrated earlier[10,11]. The aim of our present study is to investigate the feasibility of *RE*NiO$_3$/*RE*'NiO$_3$ thin film multilayers. Rare earth nickelates have in the past attracted interest because of their sharp metal-to-insulator (MI) transition whereof the critical temperature $T_{MI}$ can be tuned with the rare earth size[13,14]. Further to the MI transition, most *RE*-nickelates exhibit in the insulating phase a complex anti-ferromagnetic ordering below the Néel temperature $T_N$.[13,14] Depending on the *RE*, these two transitions occur either at the same or at a distinct temperature.[12-15] Among the different possible applications, the control of the conductivity or even of the MI transition by an external electric field is of particular technological interest with the potential for novel switches or sensors, as suggested by recent reports on ultrathin LaNiO$_3$ or NdNiO$_3$ films.[16,17] Finally, interest in nickelates stems also from recent reports regarding potential multiferroic properties [14,18-20] and oxygen sensor devices.[21]

Generally speaking, the most stable multilayer structures are those that have low in-plane lattice mismatch and for our study we have chosen to deposit SmNiO$_3$/NdNiO$_3$ (SNO/NNO) multilayers which have close pseudo-cubic lattice parameters ($a_{pc-SNO}$ = 3.796 Å, $a_{pc-NNO}$ = 3.807 Å ). These two nickelates are also interesting to combine because for NdNiO$_3$ the magnetic and MI transitions occur at the same temperature (T$_{MI}$ = T$_N$ ≈ 200 K) while they are distinct for SmNiO$_3$ (T$_{MI}$ ≈ 403 K, T$_N$ ≈ 225 K) and this although the ionic radii of Nd$^{3+}$ and Sm$^{3+}$ differ by only 0.03 Å.[22]

[(SmNiO$_3$)$_{t1}$/(NdNiO$_3$)$_{t2}$]$_n$ thin film multilayers have been deposited on polished single crystalline (001) oriented LaAlO$_3$ (LAO) substrates because of the good lattice mismatch ($a_{pc-LAO}$ = 3.791 Å). Both bilayers with $n$ = 1 and multilayers with 1 < $n$ ≤ 40 were obtained by injection Metal-Organic Chemical Vapour Deposition (MOCVD) "band flash" using 2,2,6,6-tetramethylheptanedionato-chelates of corresponding metals as volatile precursors. More detailed deposition conditions can be found in ref. [23].

The chemical composition has been checked for the thickest multilayers by Wavelength Dispersive Spectroscopy (WDS) using a CAMECA SX50 spectrometer. The atomic ratio between the three cations is stable between the different multilayers with a value Sm / Nd / Ni = 0.55 / 0.45 / 1, which suggests that the thicknesses $t_1$ and $t_2$ of SmNiO$_3$ and NdNiO$_3$ within a given multilayer are not strictly the same. The total thickness of the samples has been determined by X-ray reflectometry (XRR) and was also confirmed by the simulation of the WDS spectra recorded at different acceleration voltages.

Figure 1 shows X-ray diffraction (XRD) data for various [(SmNiO$_3$)$_t$/(NdNiO$_3$)$_t$]$_n$ multilayers deposited on LaAlO$_3$ , where $t$ is the expected thickness in Å from the deposition conditions. The only reflections obtained for the SNO/NNO multilayers phase are 00$\ell$ (pseudo-cubic indexation) which coincide with 00$\ell$ reflections of the substrate whatever $t$ and $n$. φ-scans (not shown here) from a 4-circles diffractometer demonstrate the epitaxy between the multilayers and LaAlO$_3$. XRD shows no evidence for impurity phase in the case of bilayers ($n$=1), whereas NiO appears as a 220-textured minor secondary phase for layers with $n$>1. Based on the observation that no excess of Ni has been detected by WDS, the presence of NiO is related to the difficulty to stabilize the Sm-Nd nickelates as a pure phase above a critical thickness for which single oxides are thermodynamically equally stable than the perovskite phase. A zoom on the 004 diffraction line in Figure 1.b. shows satellite peaks for the thickest films which provide evidence for the multilayer structure and allow the determination of the thickness of the SmNiO$_3$/NdNiO$_3$ bilayer for the n=25 and n=100 multilayers. Table 1 summarizes the different thicknesses deduced from WDS, XRR and XRD data.

**Table I**
Summary and thicknesses of $[(SmNiO_3)_{t1}/(NdNiO_3)_{t2}]_x$ multilayers investigated in this study

| Notation in this paper | Total thickness of multilayer (Å) from XRR | Thickness (Å) of $(SmNiO_3/NdNiO_3)$ bilayer from XRD | Thickness of single layers from WDS (Å) | |
|---|---|---|---|---|
| | | | $t_1$ SmNiO$_3$ | $t_2$ NdNiO$_3$ |
| $[(SmNiO_3)_{25} / (NdNiO_3)_{25}]_1$ | 40 ± 2 | | | |
| $[(SmNiO_3)_{25} / (NdNiO_3)_{25}]_7$ | 272 ± 10 | | | |
| $[(SmNiO_3)_{25} / (NdNiO_3)_{25}]_{40}$ | 1520 ± 45 | 37 ± 2 | 20 ± 1 | 17 ± 1 |
| $[(NdNiO_3)_{25} / (SmNiO_3)_{25}]_7$ | 270 ± 10 | | | |
| $[(SmNiO_3)_{100} / (NdNiO_3)_{100}]_1$ | 156 ± 5 | | | |
| $[(SmNiO_3)_{100} / (NdNiO_3)_{100}]_{10}$ | 1450 ± 45 | 169 ± 31 | 95 ± 5 | 74 ± 5 |

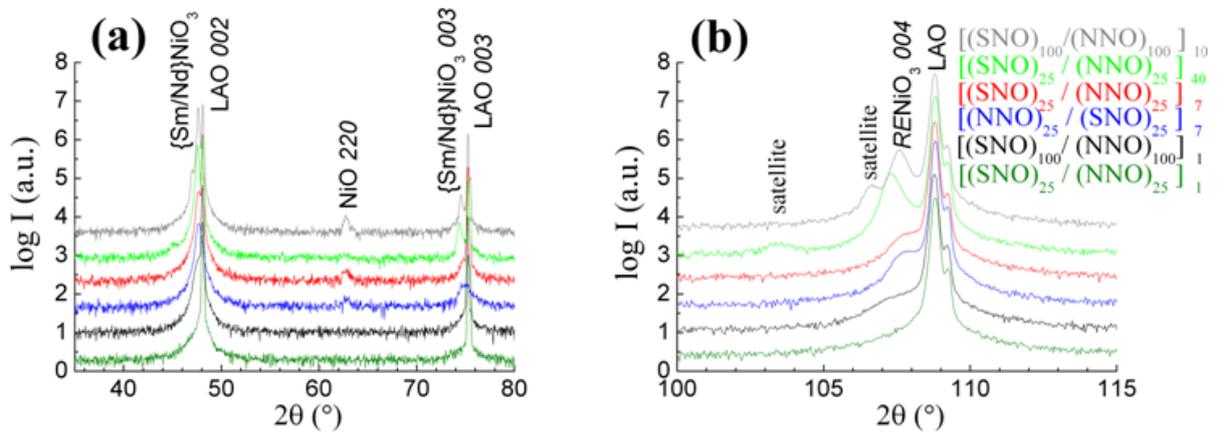

**Figure 1 (colour on-line)**
Comparison of room temperature X-ray diffraction θ/2θ scans for $[(SmNiO_3)_{t1}/(NdNiO_3)_{t2}]_x$ multilayers on LaAlO$_3$ substrates.

Although the X-ray scattering results demonstrate the stability of the $RE$NiO$_3$ perovskite structure, the resolution of our experimental set-up does not allow differentiating the very close lattice parameters of SNO and NNO, and evidently even less to discuss the potential presence of a solid solution at the interface. In order to address these questions, we have used Raman spectroscopy which is known to be a versatile technique for the investigation of thin film oxides[24-28], including perovskite-type multilayers[29,30]. The presented Raman spectra were recorded on a LabRam Jobin-Yvon spectrometer, with a 514.5-nm excitation line of an Ar$^+$ ion laser. Following earlier reports on the laser-power-dependence of Raman spectra of nickelate thin films[18,28,31], our experiments have been carried out using low powers of less than 1 mW to avoid overheating of the samples.

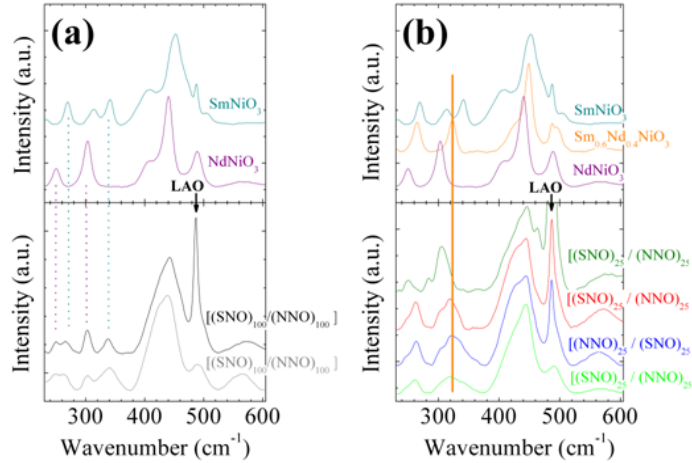

**Figure 2 (colour on-line)**
Comparison of room temperature Raman spectra for (a) [(SmNiO$_3$)$_{100}$/(NdNiO$_3$)$_{100}$]$_x$ and b) [(SmNiO$_3$)$_{25}$/(NdNiO$_3$)$_{25}$]$_x$ multilayers on LaAlO$_3$ substrates. For comparison, the top panel presents literature from ref. 18. Raman spectra for individual SmNiO$_3$ (SNO), Sm$_{0.6}$Nd$_{0.4}$NiO$_3$ (SNNO) and NdNiO$_3$ (NNO) thin films. The arrow indicates a Raman band from the LaAlO$_3$ substrate.

Figure 2.a presents Raman spectra of a [(SNO)$_{100}$/(NNO)$_{100}$]$_1$ bilayer and a [(SNO)$_{100}$/(NNO)$_{100}$]$_{10}$ multilayer on a LAO substrate. For comparison earlier reported Raman reference data[18] for single layer SNO, NNO and Sm$_{0.60}$Nd$_{0.40}$NiO$_3$ (SNNO) thin films are also shown. Apart from the sharp substrate mode at 490 cm$^{-1}$ observed for the thinner bilayer, the Raman spectra of the bilayer and the multilayer are similar. A comparison with the reference data shows that the bi-/multilayer spectra can be explained by a superposition of the Raman spectra of individual SNO and NNO layers. This observation demonstrates the presence and stabilisation of both SNO and NNO nickelates in opposition to a thickness modulated Sm$_{1-x}$Nd$_x$NiO$_3$ solid solution. A closer inspection of Figure 2.a shows two further features: (i) The positions of individual modes are slightly different between both the individual layers and the multilayers and in-between the bi- and multilayer. Although the important overlap of most bands inhibits a meaningful analysis of such shifts, already this observation suggests different strain states between the different films. (ii) Most bands are slightly larger in the multilayer than in the bilayer. This signature suggests a reduced structural coherence in the thicker multilayer, which is in agreement with the expected and observed[23] overall trend of a structural degradation of strain-stabilized phases with increasing thickness.

Figure 2.b presents Raman data for three [(SNO)$_{25}$/(NNO)$_{25}$]$_x$ multilayers of which the spectral signature is at first similar to the above-discussed [(SNO)$_{100}$/(NNO)$_{100}$]$_x$ multilayers, but also presents significant differences. Namely, the two characteristic bands which are observed at 305 cm$^{-1}$ (NNO) and 340 cm$^{-1}$ (SNO) in the [(SNO)$_{100}$/(NNO)$_{100}$]$_x$ multilayers transform into a broad feature for the [(SNO)$_{25}$/(NNO)$_{25}$]$_x$ multilayers with a significantly increased intensity around 325 cm$^{-1}$ pointed by a vertical line. The consistence of this feature with the characteristic Raman band at 323 cm$^{-1}$ of the SNNO thin film strongly suggests that the [(SNO)$_{25}$/(NNO)$_{25}$]$_x$ multilayers are not only made up by SNO and NNO but furthermore present a $Sm_{1-x}Nd_xNiO_3$ solid solution. From the comparison with the SNNO Spectrum this solid solution appears to have a chemical composition close to 0.5 and is expected to be located at the interface due to interdiffusion at the interface of the SNO and NNO layers.

It is known that the identification of the perovskite phase in RE-nickelate films by XRD is not sufficient to attest the quality of thin films, i.e. films with an identical θ-2θ XRD pattern can either present or not present the characteristic MI-transition. Although a detailed physical characterisation of the multilayers is beyond the scope of our letter, we have measured the temperature-dependent electrical properties of a representative [(SNO)$_{100}$/(NNO)$_{100}$]$_1$ bilayer to further validate the structural quality of our samples. Figure 3 shows the temperature-induced evolution of the resistivity for a [(SNO)$_{100}$/(NNO)$_{100}$]$_1$ bilayer measured from 80 to 500 K using the four-probe technique; the resistivity is normalized to its minimum value reached at the metal-insulator transition temperature $T_{MI}$. For comparison, earlier reported data[18] for single layer SNO, NNO and $Sm_{0.6}Nd_{0.4}NiO_3$ (SNNO) films is also shown. The resistivity vs. temperature of the [(SNO)$_{100}$/(NNO)$_{100}$]$_1$ bilayer shows two transitions determined from the point of deflection of the resistivity curve: (*i*) a first transition upon heating at $T_{MI(1)} \approx 160$ K, which is very close to the value of $T_{MI} = 158$ K for the single NNO layer[18]. Similarly to the single NNO layer, the resistivity of the bilayer presents a hysteresis indicating a first-order phase transition, although we note that the width of the bilayer hysteresis is smaller than in the individual film. (*ii*) a second transition, less pronounced and less defined, at $T_{MI(2)} \approx 370$ K which shows no hysteresis, similarly to SNO; this temperature is lower than $T_{MI} = 393$ K of a single SNO film. This lowering of the electrical transition temperature may be attributed to strain effect[30] which can occur in the bilayer due to the growth of NNO above the SNO layer; the presence of a $Sm_{1-x}Nd_xNiO_3$ solid solution at the interface is also consistent with the lowering of $T_{MI(2)}$ and the large temperature plateau between 340K-400K where the resistivity remains constant before the final metallic regime.

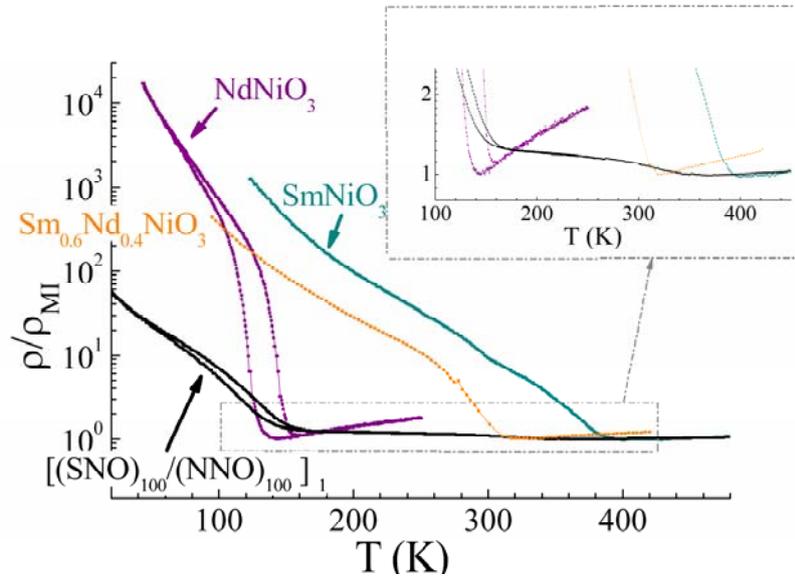

**Figure 3 (colour on-line)**
Resistivity vs. temperature measurements for a [(SmNiO$_3$)$_{100}$/(NdNiO$_3$)$_{100}$]$_1$ bilayer on LAO. For comparison literature data from ref. 18 in of single layer SmNiO$_3$, NdNiO$_3$ and Sm$_{0.6}$Nd$_{0.4}$NiO$_3$ thin films on LAO are also shown.

In summary, we have presented the synthesis, and investigation by XRD and Raman scattering of [(SmNiO$_3$)$_t$/(NdNiO$_3$)$_t$]$_n$ thin film multilayers. The multilayer structure and the presence of the metastable phases SmNiO$_3$ and NdNiO$_3$ are evidenced by X-ray and Raman scattering, thus illustrating their possible stabilization in a multilayer architecture. Electric measurements of a bilayer structure further support the presence and structural quality of the embedded rare earth nickelate layers. Future work will concentrate on the more detailed characterisation of magnetic and electric properties of multilayers. The observed feasibility of nickelate multilayers might well be extended to all rare earth nickelates giving rise to interesting coupling phenomena. We hope that this study encourages the use of other deposition techniques to still increase the quality of such nickelate multilayers and to explore other sublayer periodicities and thicknesses. Ultrathin sublayer thicknesses will be namely needed for further exploring the proposed[16,17] field effect devices based on RE-Nickelates.

*Acknowledgements*
The authors thank H. Roussel for help in x-ray characterization and J. Marcus (Institut Néel, Grenoble) is acknowledged for access to electric transport measurements.